\documentclass[a4paper]{aa}  
\usepackage{graphicx}
\usepackage{txfonts}
\usepackage{natbib}
\usepackage{multirow}
\newcommand{\lya}{\ifmmode {\rm Ly}\alpha \else Ly$\alpha$\fi}
\def\msun{\ifmmode M_{\odot} \else M$_{\odot}$\fi}
\def\zsun{\ifmmode Z_{\odot} \else Z$_{\odot}$\fi}
\def\lsun{\ifmmode L_{\odot} \else L$_{\odot}$\fi}

\begin{document}
   \title{Keck spectroscopic survey of strongly lensed galaxies in Abell 1703: further evidence for a relaxed, unimodal cluster}
   \titlerunning{Keck spectroscopic survey in Abell 1703: further evidence for a relaxed, unimodal cluster}
   %\titlerunning{Keck spectroscopic survey in Abell 1703}
   \author{
Johan Richard\inst{1,2}\fnmsep\thanks{Marie-Curie fellow}
          \and
Liuyi Pei\inst{2}
          \and
Marceau Limousin\inst{3,4}
          \and
Eric Jullo\inst{5}
         \and 
Jean-Paul Kneib\inst{5}
          }

   \offprints{J. Richard}
   \institute{
Durham University, Physics and Astronomy Department, South Road, Durham DH3 1LE, UK, Johan.Richard@durham.ac.uk
         \and
%2
Caltech Astronomy, MC105-24, Pasadena, CA 91125, USA 
         \and
Laboratoire dÕAstrophysique de Toulouse-Tarbes, Universit\'e de Toulouse, CNRS, 57 avenue dÕAzereix, 65000 Tarbes, France
       \and
Dark Cosmology Centre, Niels Bohr Institute, University of Copenhagen, Juliane Marie Vej 30, 2100 Copenhagen, Denmark
        \and
Laboratoire d'Astrophysique de Marseille, OAMP, CNRS-Universit\'e Aix-Marseille, 
38 rue Fr\'ed\'eric Joliot-Curie, 13388 Marseille Cedex 13, France
             }

   \date{Received ; accepted }

% \abstract{}{}{}{}{} 
 
  \abstract
   {
Strong gravitational lensing is a unique tool that can be used to model with great accuracy the inner mass distribution of massive galaxy clusters. 
In particular, clusters with large Einstein radii provide a wealth of multiply imaged systems in the cluster core. 
Measuring the redshift of these multiple images provide strong constraints to determine precisely the shape of the central dark matter profile.
}
   { 
This paper presents a spectroscopic survey of strongly lensed galaxies in the massive cluster lens Abell 1703, 
displaying a large Einstein radius (28\arcsec at $z=2.8$) and numerous strongly-lensed systems including 
a central {\it ring}-like configuration.
}
   {
We used the LRIS spectrograph on Keck to target multiple images and lensed galaxy 
candidates, and use the measured spectroscopic redshifts to constrain the 
mass distribution of the cluster using a parametric model. 
}
   { 
The spectroscopic data enable us to measure accurate redshifts for 7 sources at $z>2$ , all of which in 
good agreement with their photometric redshifts. We update the identification of multiply imaged 
systems by discovering 3 new systems and identifying a radial counter image. We also report the discovery 
of a remarkably bright $\sim3.6$ L$^*$ $i$-band dropout at $z=5.827$ in our mask that is only moderately magnified by the 
cluster ($\mu\sim3.0\pm0.08$). The improved parametric mass model, including 16 multiple systems with 10 spectroscopic redshifts,
 further constrain the smooth cluster-scale mass distribution  with a generalized NFW profile of best-fit logarithmic slope $\alpha=0.92\pm0.04$, 
 concentration $c_{200}=4.72\pm0.40$ and scale radius $r_s=476\pm45$ kpc. The overall RMS in the image plane is 1.3\arcsec.
 }
{
Using our strong-lensing model, we predict a large scale shear signal that is consistent with weak-lensing measurements 
inferred from Subaru data out to 4 Mpc $h^{-1}$. Together with the fact that the strong-lensing modeling requires a  single dark matter clump, 
this suggests that Abell 1703 is be a relaxed, unimodal cluster. This unique cluster could to be probed further using deep X-ray, SZ and dynamics analysis, 
for a detailed study of the physics in a relaxed cluster.
}

   \maketitle
%
%________________________________________________________________
%%%%%%%%%%%%%%%%%%%%%%%%%%%%%%%%%%%%%%%%%%%%%%%%%%%%%%%%%%%%%%%%

%%%%%%%%%%%%%%%%%%%%%%%%%%%%%%%%%%%%%%%%%%%%%%%%%%%%%%%%%%%%%%%%

\section{\label{intro} Introduction}

Massive galaxy clusters with large Einstein radii ($R_E\gtrsim 20$\arcsec) offer a wealth of information about their central 
mass distribution, by displaying large numbers of multiple images in their strong lensing regions. The most striking  
case is the cluster Abell 1689, with $R_E\sim50$\arcsec \citep{Broadhurst05, Limousin07}, which holds the largest number of 
strong lensing constraints in a single field. This gives a great opportunity to map inner mass profiles to a high level of precision. 
Parametric cluster-lens mass models have been widely used \citep{Kneib96,Smith05,Richard07}, but to construct accurate mass 
models and obtain their fiducial parameters, accurate redshift information is necessary.
Indeed, one redshift of a set of multiple images enables us to estimate the absolute total mass within the Einstein radius. 
Adding many more redshifts allow to measure the full central mass profile accurately.
 Even though photometric redshifts offer an acceptable level of precision for such modeling \citep{Broadhurst05, Marusa}, spectroscopic redshifts 
are the most reliable way to confirm the identifications and find new multiply-imaged systems. 
Lowering down the redshift error on the multiple systems with spectroscopy has another great interest, as one can use strong-lensing 
 to constrain the cosmological parameters $w$ and $\Omega_m$ based on purely geometrical considerations \citep{Link,Golse}.

On the theoretical side, \citet{Broadhurst08b} recently compared the measured lensing parameters for a sample of clusters with 
large $R_E$, with predictions from the standard $\Lambda$CDM cosmology, and report a 4-$\sigma$ 
discrepancy in the presence of such large $R_E$ for the corresponding virial mass. Their result, based on a sample of massive haloes 
derived from the Millennium simulation, includes a full treatment of the projection effects, which largely bias this sample towards alignments of the 
triaxial dark-matter haloes along the line of sight. 
This bias is also pointed out by \citet{Oguri}, who investigated a  full-sky Monte-Carlo realizations of massive clusters. They show that although 
 the population of clusters with large Einstein radii is scarce, it is possible to find Einstein radii as large as 50\arcsec\  in the whole sky, under 
 the assumptions of a WMAP5 cosmology. The main difference beween \citet{Broadhurst08b} and \citet{Oguri} is the latter's treatment of the sample 
 as a whole-sky survey, which is far more representative of the observed population than the Millennium simulation, which only covers a small volume 
 of $\sim$5000 Mpc$^3$.This releases any disagreement between the presence of large $R_E$ and $\Lambda$-CDM. 
% However this result, based on a sample of massive haloes derived from 
%the Millennium simulation, might not be representative of an all-sky survey for such extreme clusters. , who 
%investigated full-sky Monte-Carlo realizations of massive clusters, the population of clusters with large Einstein radii is scarce, and largely 
%biased towards  They find that Einstein radii as large as 50\arcsec\ can be found 
%in the whole sky with a WMAP5 cosmology, which releases any disagreement between the presence of large $R_E$ and $\Lambda$-CDM. 

%Measurements of the Einstein radius 
%There has been some discrepancy in the measurement of the concentration parameter $c_{vir}$, in particular for the case of Abell 1689 \citep{Broadhurst08a,Umetsu}, where 
%some of the weak-lensing measurements seem to disagree with the strong-lensing observations, largely underpredicting the high Einstein radius $R_E$ and measuring 
%very large concentration values \citep{Comerford}. As Abell 1689 shows a complex mass distribution, the possibility of this cluster undergoing a merger
%along the line of sight might explain the disagreement between independent strong-lensing and weak-lensing 
%measurements of the concentration \citep{Limousin07}. 

Among the clusters with large $R_E$, we have started to study in \citet{Limousin08}, hereafter L08, the mass distribution of 
Abell 1703, one of the richest clusters discovered in the Sloan Digital Sky Survey (SDSS) and showing strong gravitational lensing 
\citep{Hennawi}. In this previous work, we have identified 13 systems forming highly-magnified 
multiple images, including a central ring. Based on the spectroscopic redshift of this central feature and photometric redshifts for the 
remaining images, all the lensing systems have been well reproduced by a single NFW \citep{NFW} profile for the dark 
matter, making Abell 1703 relatively {\it simpler} compared to bimodal  lensing clusters (e.g. \citealt{Ardis} for Abell 2218, \citealt{Richard07} for Abell 68, 
\citealt{Verdugo} for MS2053). We therefore decided to  use it as a more reliable probe for comparing the results of strong and weak-lensing measurements in clusters  
with large Einstein radii from \citet{Broadhurst08a}.

We present in this paper results from a spectroscopic survey conducted at Keck using the LRIS instrument, aiming at confirming 
the photometric redshifts, multiply images identification, and precisely constraining the mass distribution in Abell 1703. We obtained
 new spectroscopic redshift measurements for most of the systems identified by L08 (10 out of 13), in good agreement with their 
 corresponding photometric redshifts. We also report of an exceptionally bright $i$-band dropout at $z\sim6$, forming a magnified 
 single image close to the region of multiple images. We further identify three new systems of multiple images and constrain 
 the lensing model, as well as compare the weak-lensing extrapolation of our model to the recently measured weak-lensing profile of 
 \citet{Broadhurst08a}.

We present in Section 2 the observations and data reduction of the spectroscopic data. Section 3 gives the redshift measurements, spectra 
and the main results from the mass modeling. We discuss these results in Section 4.
Throughout the paper, we used a standard $\Lambda$-CDM model with $\Omega_m=0.3$, $\Omega_\Lambda=0.7$ and $H_0=70$ km s$^{-1}$ 
Mpc $^{-1}$, whenever necessary. In this cosmology, 1\arcsec\ on the sky is equivalent to a physical distance of 4.244 kpc at the redshift z = 0.28 of the cluster. All magnitudes are quoted in the AB system.
%__________________________________________________________________
%%%%%%%%%%%%%%%%%%%%%%%%%%%%%%%%%%%%%%%%%%%%%%%%%%%%%%%%%%%%%%%%

%%%%%%%%%%%%%%%%%%%%%%%%%%%%%%%%%%%%%%%%%%%%%%%%%%%%%%%%%%%%%%%%

\section{\label{data}Observations}
\subsection{Imaging and photometric redshifts}

The current study is based on the same imaging data described in L08, based on Hubble Space Telescope and Subaru observations 
of Abell 1703 covering broad-band filters from $B$ to $H$. These images were previously used to derive photometric redshifts with the 
software {\tt Hyperz} \citep{hyperz}. We briefly summarize the characteristic parameters, central wavelength and depth, of these images in Table \ref{images}.

\subsection{Keck spectroscopy}

We have used the Low Resolution Imager and Spectrograph (LRIS, \citealt{lris}) on the Keck telescope to perform 
multi-slit observations of the Abell 1703 cluster field. We designed a multi-slit mask containing 32 slits of 1.0\arcsec\ width to include 
as many of the multiple systems previously identified in L08 as possible, with a few tilted slits following the geometry of 
long arcs (Fig. \ref{general}). In the case of fainter or less reliable identifications, independent images of the same system were observed in separate 
slits. Additional slits in the mask included new multiple systems candidates, cluster members and background galaxies as estimated 
through their photometric redshifts.

The LRIS instrument was set up with a 6800 \AA\ dichroic isolating the blue arm, with a 300 lines/mm grism blazed at 5000 $\AA$, 
from the red arm, with a 600 lines/mm grating blazed at 7500 \AA. This ensures a high throughput over the wavelength range 
3500 to 9500 \AA\ and good spectral resolution in the red to resolve skylines as well as $[{\rm OII}]$ emission line doublets 
in the redshift range $0.85-1.40$. This instrument setup is also well-matched with the photometric redshift estimates of the targetted sources. 
We obtained 2 hours of exposure time on 2008 May 10 in photometric conditions and very good seeing (0.7''). 

\subsection{Data reduction and redshift measurements}
\label{zmeasure}
Data reduction of the spectra 
was performed using the Python version of the \citet{Kelson} reduction scripts, which offer the advantage of processing the 
images in their distorted framework. This helps to reduce noise correlations, in particular for the case of tilted slits. We 
performed standard reduction steps for bias removal, flat-field correction, wavelength calibration, sky subtraction and cosmic-ray 
rejection, and used observations of the standard star BD+33-2642 obtained on the same night to derive the flux calibration. 

Close-ups on the targets and extracted spectra are presented in Fig. \ref{spectra} and \ref{spectra2}.
We measured the redshift for 33 sources in the entire LRIS mask. The spectra show either Lyman-$\alpha$ 
(in  absorption or emission) with additional ultraviolet absorption lines in the blue part, or a resolved doublet of $[{\rm OII}]$) 
in the red part of the spectrum. In two multiple systems (10/11 and 16), we observed these 
images in multiple slits: redshift measurements were obtained by identifying significant absorption lines after stacking the relevant 
exposures. For each spectrum, the average redshift value is obtained from the peaks of the main spectral features identified, while the corresponding 
error is taken from the spectral dispersion. Additional uncertainties generated by the accuracy
of the relative and absolute wavelength calibrations, of about 1.1 and 1.5 \AA, respectively, were quadratically added to yield the final redshift
errors. A confidence class $z_{class}$, ranging from 1 to 4, was assigned to
each measurement according to the prescription
of \citet{lefevre}: this corresponds to a probability
level for a correct identification of 50\%, 75\%, 95\%, and 100\%,
respectively. A specific value of 9 is used when only a single secure
spectral feature is seen in emission.

Table \ref{zspec} summarizes the redshift measurements $z_{spec}$, confidence class $z_{class}$ and the spectral features used to derive the redshift for all 
objects detected in the LRIS mask, including those additional targets serendipitously falling into the slits. For the majority of sources located within the ACS 
field of view, we report in the same table the corresponding photometric redshift $z_{phot}$. We find a good agreement
between individual $z_{spec}$ and $z_{phot}$ values, with a typical error $<|z_{phot}-z_{spec}|>= 0.15$ 
or $<|\frac{z_{phot}-z_{spec}}{1+z_{spec}}|>$=0.047, and no {\it catastrophic} measurements. This strong consistency is mainly 
owing to the use of space-based images of very similar PSF size, together with ground based near-infrared images obtained with MOIRCS in excellent seeing 
conditions (0.4\arcsec ). 

%__________________________________________________________________
%%%%%%%%%%%%%%%%%%%%%%%%%%%%%%%%%%%%%%%%%%%%%%%%%%%%%%%%%%%%%%%%

%%%%%%%%%%%%%%%%%%%%%%%%%%%%%%%%%%%%%%%%%%%%%%%%%%%%%%%%%%%%%%%%

\section{Results}

\subsection{Multiple images redshifts and identifications}

\begin{figure*}
\centerline{\mbox{\includegraphics[height=15cm,angle=0]{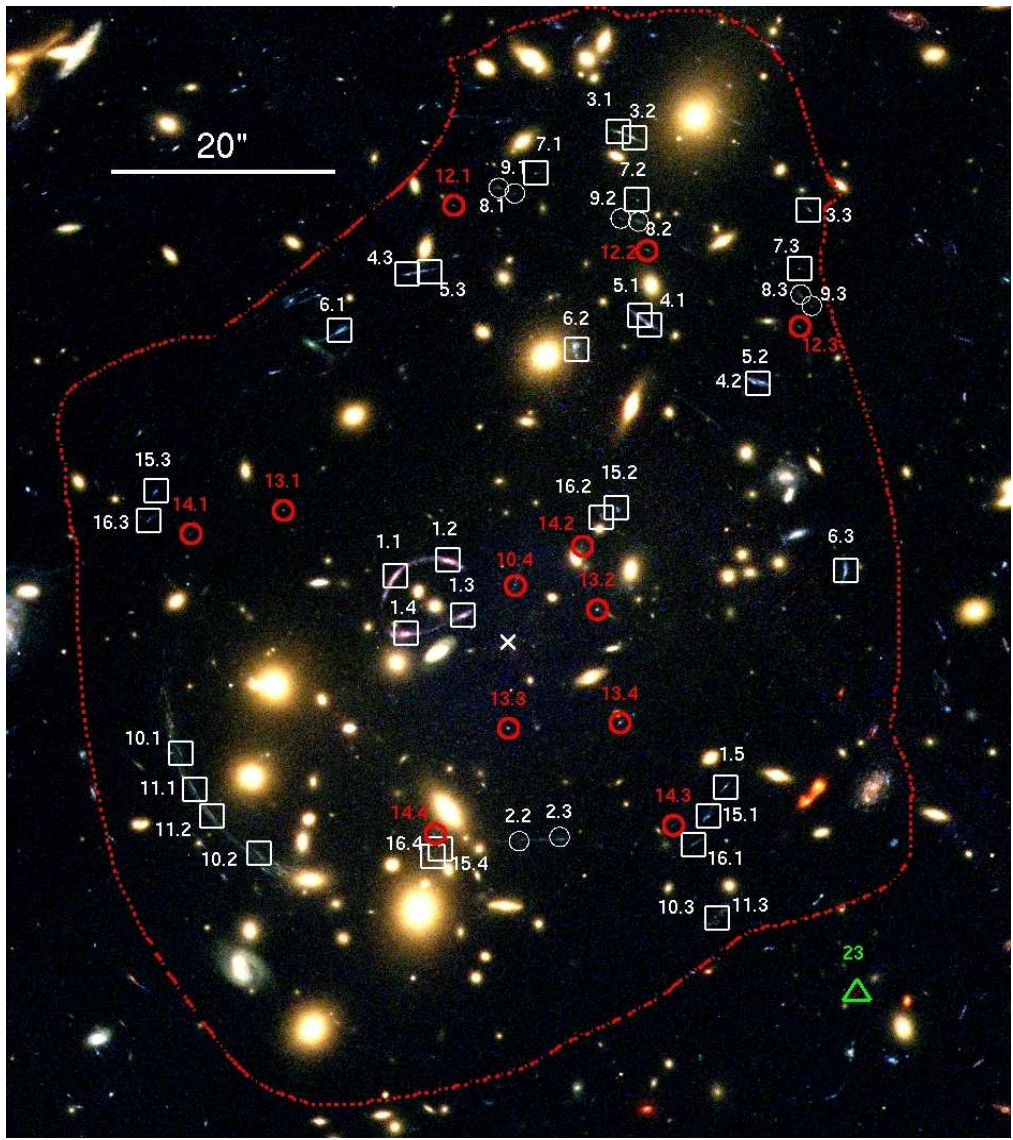}}}
\caption{\label{general}ACS color image of Abell 1703 (combination of the HST filters F450W, F606W and F850LP), showing the location 
of all the multiply-imaged systems included in this study. 
The white cross at the center of the image marks the location of the brightest cluster galaxy, which has been subtracted for clarity.
The dashed line outlines the limit of the region where we expect multiple images 
up to $z=6$. The systems presented in L08 are shown in white, and the new systems presented in this work are shown in red. Squares identify  
the multiple images with spectroscopic redshifts.The green triangle located to the lower right 
marks the location of the $i$-band dropout presented in Section \ref{dropout}.}
\end{figure*}

Figure \ref{general} presents a color image of the central region of Abell 1703, combining the $B$, $V$ and $Z$ ACS filters, where we 
marked in white the location of the 11 multiply imaged sources identified in our previous work (L08).
Our spectroscopic data enabled to measure the redshifts for the following 8 systems (displayed as white squares in Fig. \ref{general}):

\begin{itemize}
\item{System 1 at $z=0.8889$ forms a very highly magnified ring-shaped configuration with 4 images close to the BCG and a less-magnified 5th image to the south-west.
 Although its redshift was previously derived in L08, we obtained a new spectrum with the same mask in much better conditions.}
\item{System 3 at $z=3.277$ is located to the north of the BCG, with 3 images: 2 bright merging images and a fainter counter-image to the west.}
\item{Systems 4/5 at $z=1.9082$ are two individual regions identified in the same source which forms a cusp configuration of 3 images.}
\item{System 6 at $z=2.360$ is a similar configuration as Systems 4/5, but closer to the center of the cluster.}
\item{System 7 at $z=2.983$ is again a cusp configuration like Systems 4/5, but further from the center of the cluster.}
\item{Systems 10/11 at $z=2.627$ are two individual regions identified in the giant tangential arc at the south-east of the BCG. A fainter counter-image 
is located to the west. The new mass model presented here enable us to associate the radial feature reported in L08, very close to the BCG (next to the ring 
configuration of System 1) to be a 4th image (10.4) associated with this arc.}
\item{System 15 at $z=2.355$ is a typical \textit{Einstein-cross} configuration of 4 images surrounding the BCG. We note that systems 6 ($z=2.360$) and 
15 ($z=2.355$) have similar redshifts and are therefore very close to one another in the source plane, with a projected separation of 67$\pm$6 kpc.
}
\item{System 16 at $z=2.810$ is another \textit{Einstein-cross} configuration located at a slightly higher redshift, but found very close
to System 15 in the image plane due to projection effects.}
\end{itemize}

Additional sources 20 and 21, which were included in the mask as a possible new multiple system, show a very similar spectroscopic redshift 
$z=1.279$ with $[{\rm OII}]$ doublet of emission lines. However, they are located too far from the center of the cluster to be multiply imaged 
at this redshift.

Apart from system 1, all the other systems are located in the redshift range $1.9<z<3.3$. 
Using the new spectroscopic redshifts into the mass model (see Sect. \ref{sl}), we have computed the predicted region of multiple images for  a  
redshift $z=6$ (delimited with a dashed line in Fig. \ref{general}). As we expect all the multiple images present in the HST optical bands to lie within 
this region, we use it to update our systematic search for multiple images based on the photometric redshifts catalog. 
We find three additional systems complementing the list presented in L08.  We display them as red circles in Fig. \ref{general} and report the corresponding positions 
and photometric redshifts in Table \ref{newmul}:

\begin{itemize}
\item{System 12 is again a typical cusp configuration of 3 images located to the north, and ``following'' the nearby Systems 4/5, 7, 8 and 9.   }
\item{System 13 and 14 are two further Einstein-cross configurations of 4 images surrounding the BCG. System 14 is located near the Systems 15 and 16 in 
the image plane, while System 13 is closer to the cluster center. }
\end{itemize}

\begin{figure*}
\caption{\label{spectra} BVI ACS color image close-up on individual $z>0.5$ sources in the LRIS mask with the location of the slit, and corresponding extracted spectra. 
The $i$-band dropout (23, see Sect. \ref{dropout}) is presented separately in Fig. \ref{23}.}
\input{newsummary1}
\end{figure*}

\begin{figure*}
\caption{\label{spectra2} Continuation of fig. \ref{spectra}}
\input{newsummary2}
\end{figure*}

\subsection{Spectroscopy of a remarkably bright $i$-dropout}
\label{dropout}

\begin{figure}[ht]
\centerline{\mbox{\includegraphics[width=8cm]{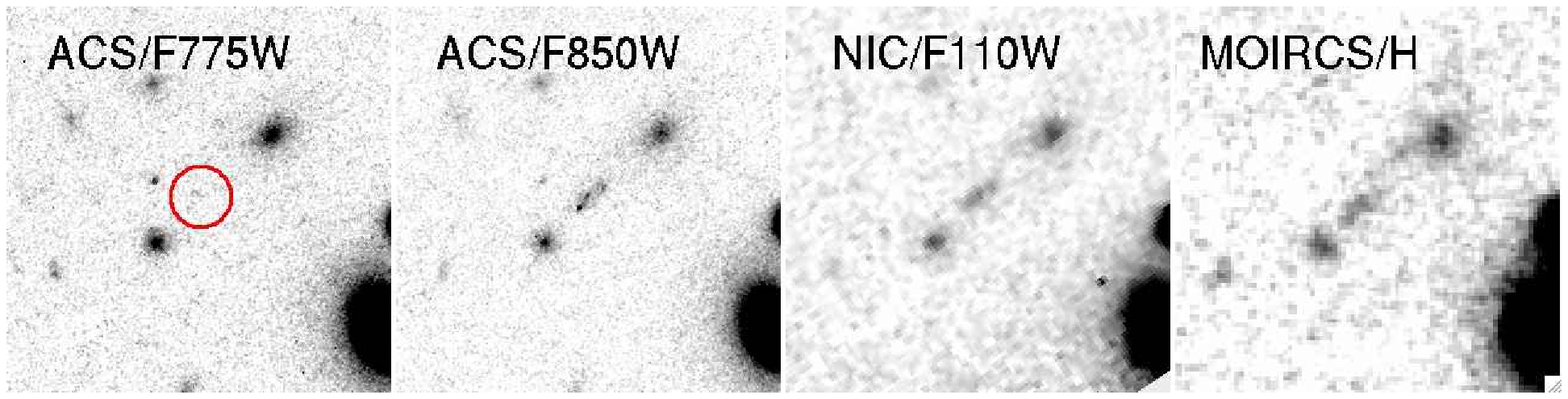}}}
\centerline{\mbox{\includegraphics[width=8cm]{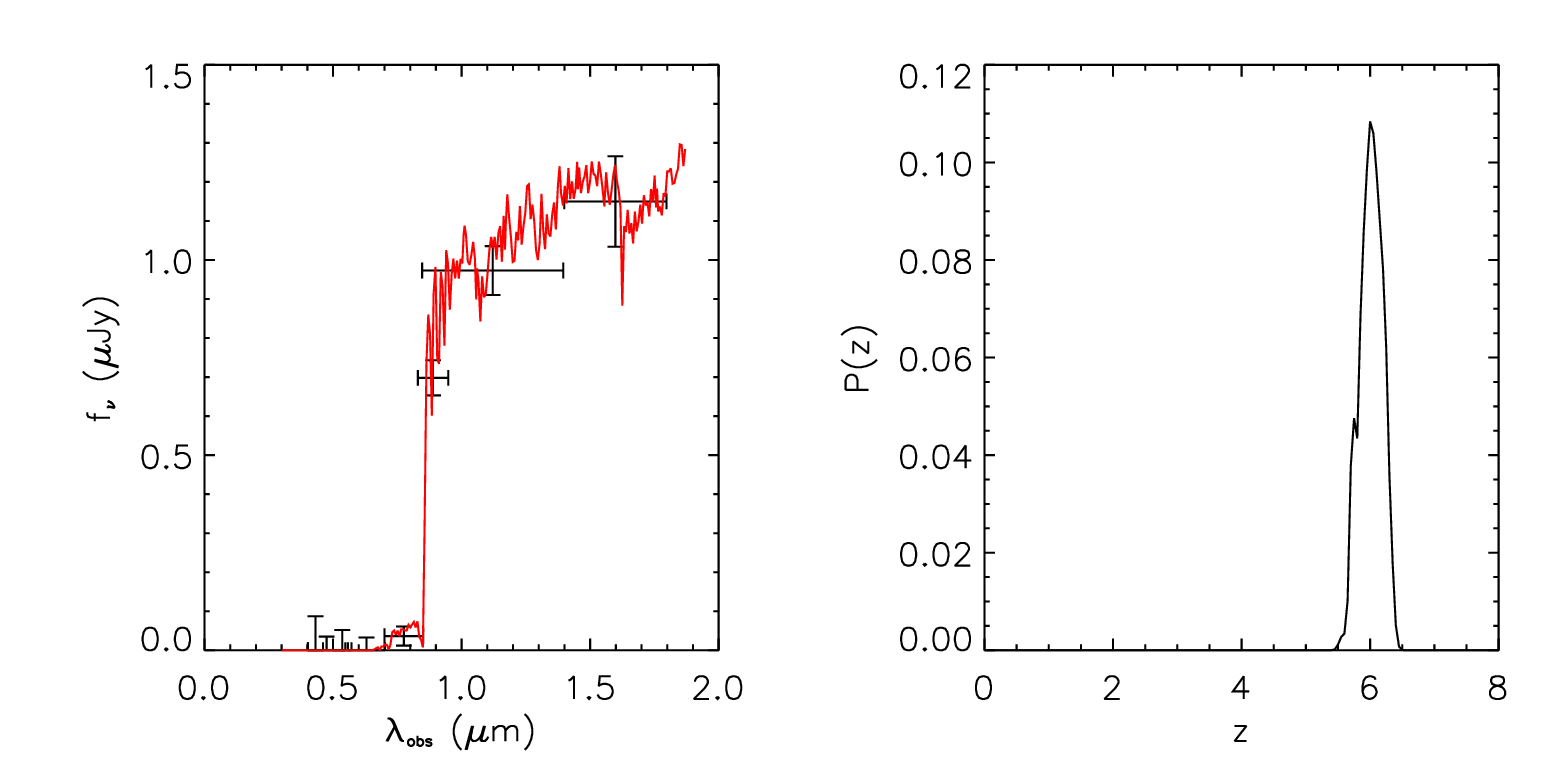}}}
\begin{minipage}{3.cm}
\centerline{\mbox{\includegraphics[width=0.95\textwidth]{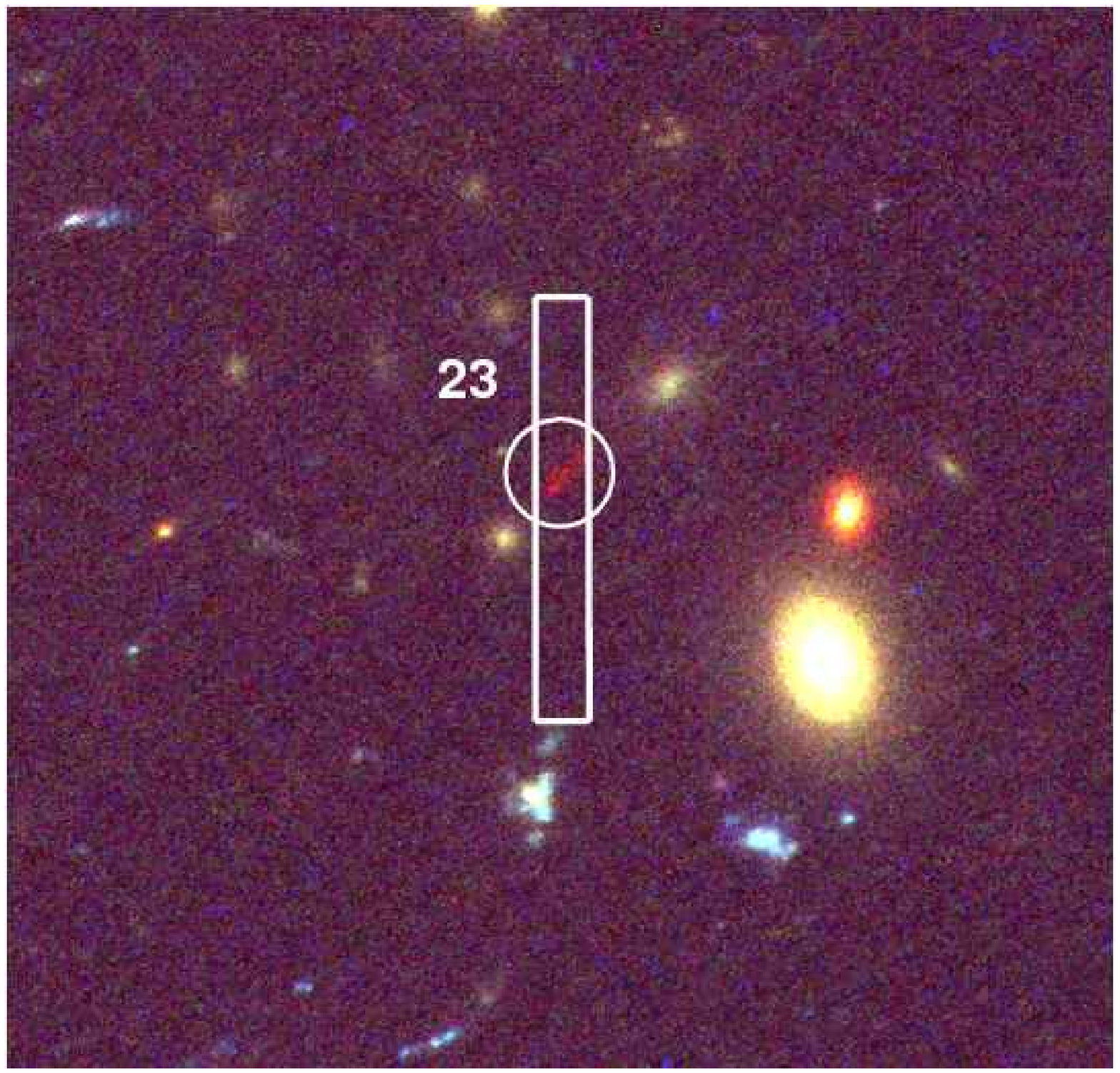}}}
\end{minipage}
\begin{minipage}{6.cm}
\centerline{\mbox{\includegraphics[width=0.95\textwidth]{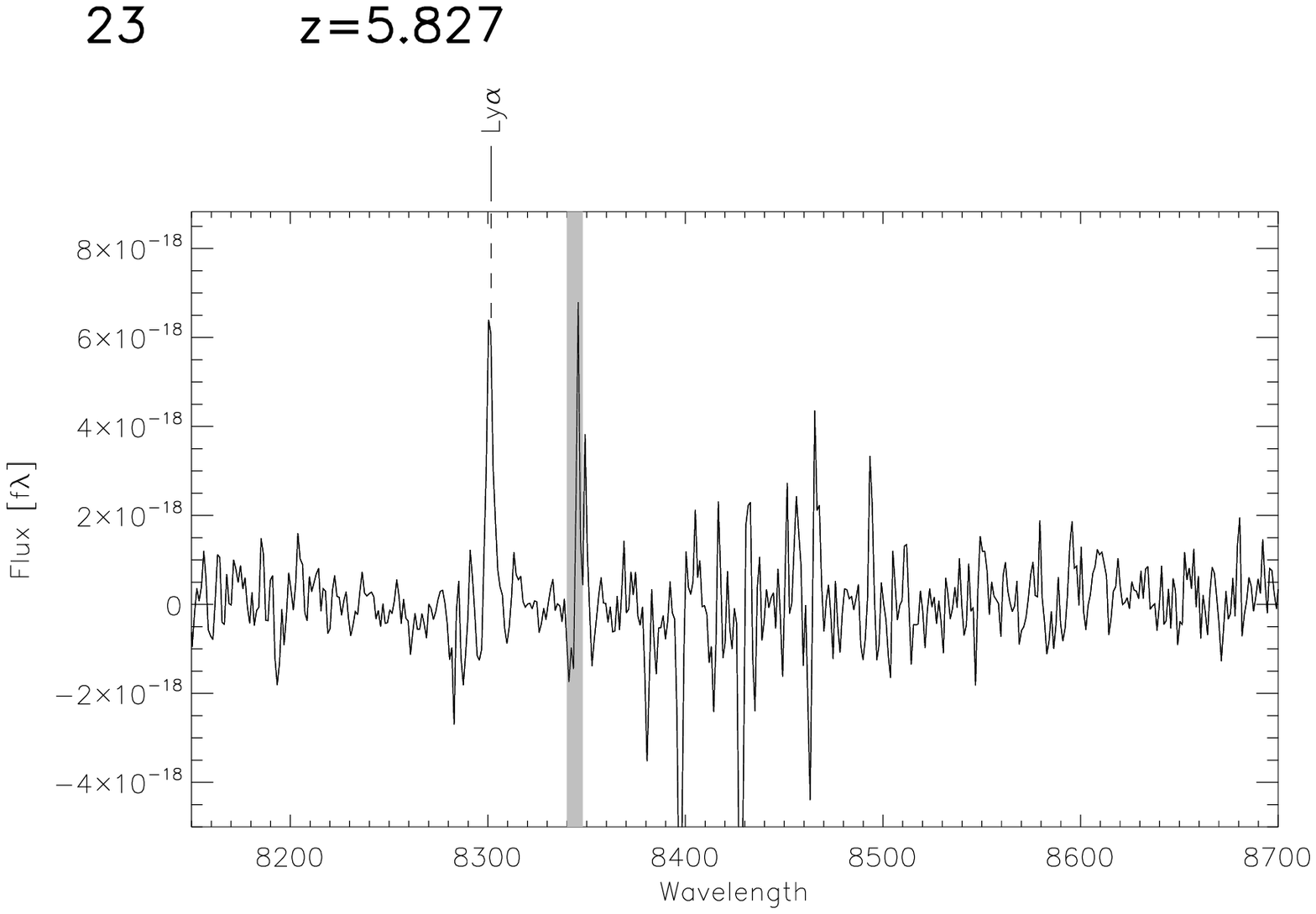}}}
\end{minipage}
\caption{\label{23} A bright $i$-dropout found in the field of Abell 1703. The top panel shows the individual detections 
in the ACS, NICMOS and MOIRCS image. The object is resolved and elongated along the shear direction, with 
an angular size 0.45\arcsec . Middle panels: best fit of the spectral energy distribution with the photometric 
redshift software (left) , and corresponding redshift probability distribution (right). Bottom panels: VIZ ACS color 
image with the location of the LRIS slit (left), and close-up on the extracted spectrum where we detected an 
emission line at 8301 \AA\ (right). The grey region highlights the location of a bright OH skyline.}
\end{figure}

One of the sources in the LRIS mask, identified as 23 in the spectroscopic catalog (Table \ref{zspec}), deserves particular attention.
Located to the south-west of the cluster center (green triangle in Fig. \ref{general}), it
has been selected for its very red $i-z=2.9\pm0.3$ color and bright $z=24.30\pm0.07$, while being undetected blueward of the $i$ band 
in the other deep ACS images. The NICMOS and MOIRCS photometry $J=23.93\pm0.07$ 
and $H=23.75\pm0.10$ further constrain its spectral energy distribution to be compatible with a photometric redshift 
$z_{phot}=6.0\pm0.42$ (Figure \ref{23}, top and middle panels), and the $i-z$ 
color make it a very strong $i$-dropout photometric candidate according to the \citet{Bouwens06} 
selection criterion.

The LRIS spectrum for this object covers the wavelength range 3800-9230 \AA , 
corresponding to $2.12<z<6.59$ for Lyman-$\alpha$. We detect a 5$\sigma$-significant emission line 
at $\lambda=8300.5$ \AA\ (Figure \ref{23}, lower panel), with an integrated flux $f=2.5\pm0.4\times 
10^{-17}$ ergs s$^{-1}$ cm$^{-2}$, and an observed equivalent width W$>50$ \AA\ when assuming a 5$\sigma$ upper limit on the underlying 
continuum.This line is slightly resolved in the spectral direction, with a 
possibly asymmetric profile. In order to estimate this asymmetry, we use the \textit{weighted skewness} 
statistics $S_w$, which is based on the third momentum of the flux distribution $f_i$ (seen as an array of size N) and 
has been used as a selection criterion for Lyman-$\alpha$ emitters (LAEs, \citealt{Kashikawa}). 
The value of $S_w$, measured in Angstroms, is defined as:

\begin{equation}
S_w=\Big(\frac{1}{I\sigma^3}\ \sum_{1}^{N} (x_i-\bar{x})^3\ f_i\Big)\ \Big(\lambda_{10,r}-\lambda_{10,b}\Big)
\end{equation}
where $(x_i)$ and $(f_i)$ are the arrays of coordinates and fluxes, of size N, $I=\sum_{i}^{N}\ f_i$, and ($\bar{x},\sigma$)
represent the mean and standard deviation of the array ($x_i$), respectively. The weight factors $\lambda_{10,r}$ and $\lambda_{10,b}$ correspond to the 
wavelengths where the flux drops to 10\% of its peak value on the red and blue sides of the line emission (see \citealt{Kashikawa} for a more detailed 
discussion about the use of $S_w$). In our case, we measure $\lambda_{10,r}=8305$ \AA\ and $\lambda_{10,b}=8298$ \AA\ on the spectrum of the dropout, and 
a corresponding $S_w=5.6\pm1.4$ \AA\ . The error on $S_w$ has been measured using a bootstrap resampling of the fluxes. The skewness value 
is slightly superior to the critical value $S_w=3$ \AA\ which has been used as a lower limit for selecting single line emitters as LAEs, in \citet{Kashikawa} and 
\citet{Shimasaku}.
The single emission line is therefore likely to be Lyman-$\alpha$ at a redshift $z=5.827$, quite  compatible with the photometric redshift prediction.

At his redshift, we compute a magnification factor  and associated error $\mu=1.2\pm0.03$ magnitudes, 
derived with the improved strong lensing model (see Sect. \ref{muerror}). We note that although this object is a single image, it is located 
very close to the region of strong lensing, allowing an accurate estimate of the magnification with our model.

After correcting for this magnification, the source is both magnified and intrinsically
bright ($H_{unlensed}=24.9$), as there are only 9 of such sources at $z\sim6$ brighter than 
$z=25.5$ AB found in the UDF, HUDF and GOODS fieds \citep{Bouwens06}. This object is similar in redshift and intrinsic magnitude 
to the spectroscopically confirmed $i$-dropouts by \citet{Stanway04} at $z=5.78$ and $z=5.83$, or the slightly magnified object at $z=5.515$ 
 found in the cluster RDCS1252.9-292 \citep{Dow}. 
Based on the latest constraints of the luminosity function at $z\sim6$ \citep{Bouwens08}, the unlensed magnitude 
corresponds to an 3.6 L$^*$ galaxy. We also note that the very red $i-z=2.9$ color is much redder than for any of the 9 dropouts with $z<25.5$ from \citet{Bouwens06}, 
detected in both the $i$ and $z$ bands, and where the reddest color was found to be $i-z=2.4$. Thanks to the magnification and the combination of filters at 
this redshift, the $i-z$ color allows to measure the average  depression factor $<D>=<1-f_{\rm obs}/f_{\rm int}>$ at $z\sim6$, between the observed and intrinsic fluxes shortward of 
Lyman-$\alpha$, due to line blanketting. The corresponding value $<D>\sim0.96$ is a lower limit, due to possible contamination of the red side of the line in the $i$ 
band filter. This value is close (but slightly superior) to the predicted value at $z=6$ ($<D>=0.92$) when following the \citet{Madau} prescription.

Although the current photometry, probing only the rest-frame UV at $z=6$, does not allow to derive physical parameters 
on this source (such as age, reddening, stellar mass), we can measure the UV spectral slope $\beta$, defined as $f_\lambda\propto\lambda^{\beta}$. Using 
the $J-H=0.18$ color, we derive $\beta\sim -1.5$, a value redder than a spectrum flat in $f_\nu$ ($\beta=-2.0$), as can be seen in Fig. \ref{23}. Although this 
value is typical of Lyman-break galaxies at $z\sim3$ ($\beta=-1.5\pm0.4$, \citealt{Adelberger}), it is redder than for the sample of 27 $i$-dropouts studied  by \citet{Stanway05}, 
who derived $\beta\sim-2.0$ based on $J$ and $H$ photometry, or the sample of lensed $z$-band dropouts presented in \citet{Richard08}. 
The use of IRAC observations probing the longer wavelengths would allow us to understand whether this 
redder color is due to an old stellar population or an effect of reddening, similar to the work done by \citet{Egami} on a highly lensed galaxy at $z\sim6.8$ discovered in Abell 2218 
\citep{Kneib04}.  

\subsection{Strong Lensing}
\label{sl}

\subsubsection {Modeling method}

Our starting point in modeling of the dark matter distribution in Abell 1703 is the mass model presented in details by L08, where 
we used the {\tt LENSTOOL}\footnote{\tt http://www.oamp.fr/cosmology/lenstool/} public software \citep{Jullo} to constrain a parametric mass model with the identified multiple systems. 
The model is optimized through the new Bayesian Markov chain Monte-Carlo (hereafter MCMC) sampler, described in detail in \citet{Jullo}. This process 
uses the observational constraints (positions of the multiply imaged systems) to optimize the parameters describing the mass distribution by matching the location of each 
image of a given system in the source plane.

The contribution of the stellar mass from the central galaxy is parametrized by a Pseudo Isothermal Elliptical Mass Distribution(PIEMD) 
using the profile derived from the photometry. The PIEMD profile, which has been widely used for the modeling of cluster-scale \citep{Smith05} 
and galaxy-scale \citep{Natarajan98} haloes, assumes each dark matter clump can be parametrized
by a central position, ellipticity $e$\footnote{We use here the definition $e=1-b/a$ of the ellipticity,where $a$ and $b$ are the semi-major and semi-minor axis of the ellipse, respectively. }
, position angle $\theta$, central velocity dispersion 
$\sigma_0$, and two characteristic radii: a core and a cut radius. The total mass is proportional to r$_{\rm cut}\sigma^2_0$.

Using this parametrization, the PIEMD mass density $\Sigma(x,y)$ takes the form:

\begin{equation}
\Sigma(x,y)=\frac{\sigma_0^2}{2G}\ \frac{r_{\rm cut}}{r_{\rm cut}-r_{\rm core}}\Big[\frac{1}{(r_{\rm core}^2+\rho^2)^{1/2}}-\frac{1}{(r_{\rm cut}^2+\rho^2)^{1/2}}\Big]
\label{elleq}
\end{equation}

where $x$ and $y$ coordinates are oriented along the position angle $\theta$, 
$\rho^2=[(x-x_c)/(1+\epsilon)]^2+[(y-y_c)/(1-\epsilon)]^2$, $\epsilon=e/(2-e)$, and $(x_c,y_c)$ is the center of the mass distribution (see also \citet{Smith05} for more details).

The cluster-scale dark-matter component is modeled with a generalized NFW (hereafter gNFW) profile 
\citep{Zhao} including a central logarithmic slope $\alpha$:

\begin{equation}
\rho(r)=\frac{\rho_c\delta_c}{(r/r_s)^\alpha(1+(r/r_s))^{(3-\alpha)}}
\end{equation}

where $r_s$ is a scale radius, $\rho_c$ the critical density and $\delta_c$ is related to the value of the concentration parameter $c_{200}$ through the 
relation:

\begin{equation}
\delta_c=\frac{200}{3}\frac{c_{200}^{3}}{ln(1+c_{200})-c_{200}/(1+c_{200})}
\end{equation}

Again, the spherical gNFW profile is generalized to an elliptical mass distribution using a relation similar to Eq. \ref{elleq}. A more detailed description 
of  the implementation of the gNFW profile in lenstool is provided by \citet{Sand}.

Individual cluster galaxies were added as small-scale PIEMD perturbers based on their photometry and shape parameters, using empirical scaling relations between their dynamical parameters (central velocity
dispersion and scale radius) and their luminosity, assuming the \citet{FJ} relation and a constant mass-to-light ratio for each galaxy (see \citealt{Covone}
for more details).The same parameters of one particular cluster galaxy to the north (852) 
were optimized individually to reproduce the configuration of the surrounding systems 3, 7, 8, 9 and 12. 

We replaced the photometric redshifts kept as free parameters in L08 by their spectroscopic equivalents for the 7 relevant sources (9 systems). 
The precision of the new model, which uses the same parametrization as L08, can be estimated using the RMS of the location of multiple 
images in the image plane, defined as :

\begin{equation}
\sigma_i=\sqrt{\sum_{j,k}({\rm xobs}_{j,k}-{\rm xpred}_{j,k})^2+({\rm yobs}_{j,k}-{\rm ypred}_{j,k})^2}
\end{equation}

where (xobs$_{j,k}$,yobs$_{j,k}$) and (xpred$_{j,k}$,ypred$_{j,k}$) are the observed and predicted location of image $j$ in system $k$, respectively. 
Using the exact same parametrization, we find a value of $\sigma_i=1.3$\arcsec similar to the result obtained in L08 (1.4\arcsec). This precision is typical 
of strong lensing works using a similar number of multiples images \citep{Richard07,Ardis}, and smaller by a factor of $\sim2$ from the one obtained for 
Abell 1689 (2.87\arcsec, \citealt{Limousin07}).

We summarize the best fit parameters of the new model in Table \ref{model}.
The main gNFW component shows a concentration $c_{200}=4.72^{+0.43}_{-0.45}$, a scale radius $r_{s}=476.9^{+51.6}_{-42.8}$ kpc, and an 
inner slope $\alpha=0.92^{+0.05}_{-0.04}$. The geometrical parameters (center, orientation and ellipticity) are very similar to the ones measured on the cD 
galaxy (see also Sect. \ref{resec} below). In comparison with the results from L08,  the strongest variations in the best fit parameters are found in the gNFW 
profile, which appears to be more concentrated (larger $c_{200}$ and smaller scale radius $r_s$), but both results are compatible at the 3$\sigma$ level. 
The addition of the new constraints (spectroscopic redshifts, new multiple systems) while keeping the same model parametrization, did not reduce the uncertainty 
in the best fit parameters, even if both models reproduce the constraints on the multiple images with a similar precision. This argues for a remaining degeneracy 
within the gNFW parameters, which cannot be completely disentangled by a mass model purely based on strong lensing.

\subsubsection{Redshift predictions and magnifications}
\label{muerror}
One of the main benefits of the Bayesian approach is that the MCMC optimization provides a large number of models 
which sample the posterior probability density function of all the parameters \citep{Jullo}. We can use these different realizations to estimate 
the average value and associated error for any given parameter or combination of parameters. 
In our case, we estimate the redshift and associated error for each of the 
new multiple systems 12, 13 and 14. These values are reported in Table \ref{newmul}.  We also 
computed the magnification factors $\mu$ and associated error for every source in our LRIS mask (Table \ref{zspec}) 
as well as the other multiple systems (Table \ref{newmul}). The magnification factors range from 0.01-0.05 mags for single images 
to about 4.8 magnitudes for the source 1.1.

\subsubsection{Weak-lensing predictions}

Our mass model, solely based on strong lensing constraints out to 50\arcsec (the region of multiple images), can be used to predict 
the tangential {\it shear} induced on background galaxies up to larger scales. The main observable with weak-lensing is the {\it reduced 
shear} $g=\gamma/(1-\kappa)$ where $\gamma$ is the gravitational shear and $\kappa$ the convergence. In order to compare our predictions with 
weak-lensing observations of Abell 1703 with Subaru described by \citet{Broadhurst08a}, we adopt the definion of the {\it tangential reduced shear} 
$g_+$ from \citet{Umetsu}, i.e. the projection of the shear perpendicularly to the direction of the cluster center (see also \citealt{Medezinski}).

\begin{figure}
\centerline{\mbox{\includegraphics[height=8cm,angle=270]{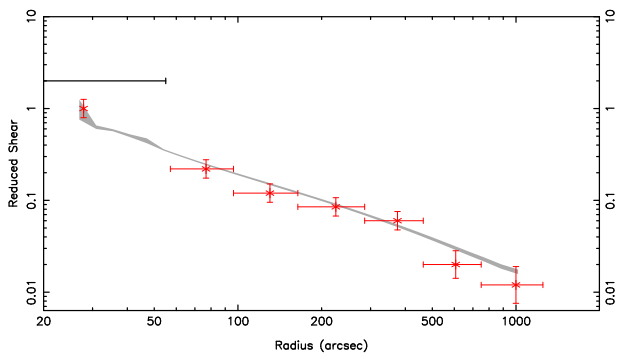}}}
\caption{\label{shear} Tangential reduced shear $g_+$ predicted with the strong lensing model based on constraints in the central region (black error bar on top). 
The mean value and 3$\sigma$ error is presented as a function of radius in the grey-filled region. We overplot the datapoints obtained by \citet{Broadhurst08a} in their 
weak-lensing measurements with Subaru.  Extrapolation of our strong lensing model forms a very good fit to these observations upto $r=1000$ arcsec, or 4 Mpc$\ h^{-1}$.}
\end{figure}

Using the {\tt MCMC} realizations, we determine predictions for $g^+$ and its associated 3-$\sigma$ error as a function of  angular distance $r$ from the 
center of the mass distribution, assuming a distribution of background sources at a fixed redshift $z_s=1.2$ corresponding the the mean of the redshift 
distribution of sources in the observations from \citet{Broadhurst08a}. Figure \ref{shear} presents these estimations 
against their weak-lensing measurements over the same range of $r$. 
We find that our strong-lensing model represents a very good fit to the weak-lensing datapoints up to 4 Mpc$h^{-1}$, with a $\chi^2$ of 4.1 / 7 = 0.58, even after 
extrapolating our mass model to a radial distance 20$\times$ larger than the region of strong lensing constraints. This argues again for the overall simplicity 
of the cluster, with a single NFW profile providing a good fit both for the strong-lensing and weak-lensing measurements.

\subsubsection{Ellipticity of the mass distribution and Einstein radius}
\label{resec}
From the best fit geometrical parameters found during the optimization of the mass model (Table \ref{model}), we notice a good 
agreement between the values for the ellipticity and position angle of the large scale {\tt gNFW} profile ($e_{\rm NFW}=0.23$, $\theta_{\rm NFW}=63.9$ degrees) and the 
same parameters of the cD galaxy ($e_{cD}=0.19$, $\theta_{CD}=52$ degrees) as measured with SExtractor
\footnote{The value of $e_{NFW}$ given in L08 had an error in its definition and we update it to the correct one.}
.This is not really surprising, as this correlation has previously been reported both in strong-lensing analysis (e.g. \citealt{Gavazzi} in MS2137), X-ray 
observations \citep{Hashimoto} as well as numerical works \citep{Dubinski,Faltenbacher} .

\begin{figure}
\centerline{\mbox{\includegraphics[height=8cm,angle=0]{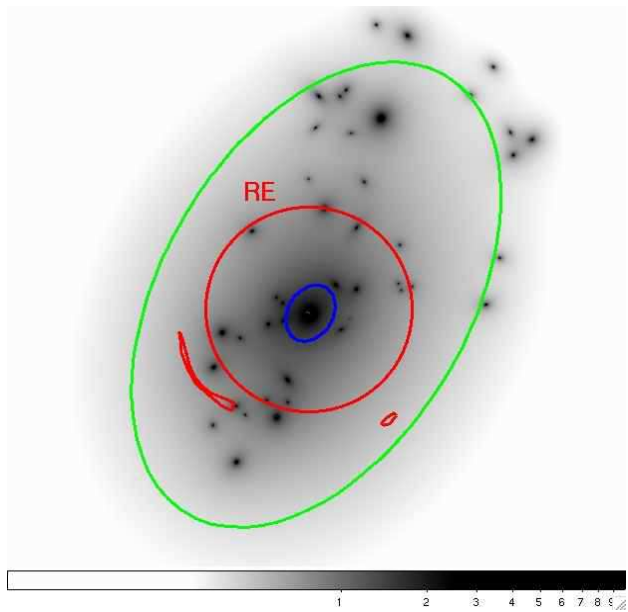}}}
\caption{\label{re} Convergence map distribution $\kappa$ for a source at $z=2.627$, the redshift of the giant tangential arc (systems 10 and 11, 
located in red). By averaging this map inside increasing apertures, we measure the corresponding {\it effective Einstein radius} $R_E$, slightly 
shorter than the distance to the giant arc due to the overall ellipticity of the mass distribution. The blue and green ellipses show the best-fit ellipses found 
on the overall mass distribution, with an increasing ellipticity when moving from small to large radii (see Sect. \ref{resec} for details).}
\end{figure}

More generally, the total mass distribution of dark matter is the sum of this {\tt gNFW} component and the various PIEMD  profiles associated with the cluster
galaxies, and the global ellipticity $e_{mass}$ may have a different value. We use the convergence map $\kappa$ (which shares an identical geometry 
with the mass distribution) to measure this ellipticity $e_{mass}$ {\it a posteriori} by fitting elliptical contours with the IRAF routine {\tt ellipse}. We keep 
the center of the ellipses fixed at the peak of the mass map and let the ellipticity $e_{mass}(a)$ and position angle $\theta(a)$ vary as free parameters 
with the semi major axis $a$. In the central region ($a<15$\arcsec), we find a best fit value $e_{mass}\sim0.23$ (Fig. \ref{re}, in blue) dominated by the 
geometry of the cD and the {\tt gNFW} profile. At larger distances $e_{mass}(a)$ increases to reach a maximum of 0.4 at $a\sim 40$\arcsec (Fig. \ref{re}, in 
green), due to an alignment of bright galaxies (as seen in projection) to the north-west and the south-east, therefore along a similar direction as $\theta_{\rm NFW}$. 
These galaxies contribute to increasing the value of $e_{mass}$ compared to $e_{NFW}$, this effect being significant on the mass distribution 
up to the edge of the ACS field of view. The same ellipticity is seen on the weak-lensing reconstructions performed by \citet{Broadhurst08a}.

We also compute the \textit{effective} Einstein Radius $R_E$, defined as the radius $r$
from the center of the cluster (located at the peak of the mass distribution) at which $\bar{\kappa}(r<R_E)=1$ \citep{Broadhurst08b}. 
We measure the value and error on $R_E$ for the redshift $z=2.627$ of the giant tangential arc (formed with the systems 10 and 11), 
using the different {\tt MCMC} realizations. We find $R_E=28.0\pm0.25$ arcsec, a value smaller than one would measure based on the simple distance 
of the giant arc (Fig. \ref{re}).  This value is close but smaller than the estimate of 32$\arcsec$ 
given by \citet{Broadhurst08b}, who mentions an agreement of $R_E$ with the distance of the arc. This apparent discrepancy is due to the ellipticity of the mass 
distribution, as the giant arc is oriented along the major axis of the elliptical mass distribution. This ellipticity is now well-constrained by the large number 
of spectroscopically confirmed multiple images presented here, making the new measurement of $R_E$ more robust.

We can use the same  cumulative probability distribution $P(>R_E)$ as \citet{Broadhurst08b} to test the agreement of this new Einstein radius with the 
$\Lambda$CDM Millennium simulation. This probability is computed from the distribution of concentrations at a given virial mass, accounting for the 
various biases arising from selecting a population of  lensing clusters, as well as the projection effects. We find a higher probability P=20\% (instead of 7.9\%) 
of agreement, making Abell 1703 less discrepant than the other mentioned clusters (Abell 1689, CL0024 and RXJ1347) for its virial mass. Together with the 
results from these 3 other clusters (P=8.5\%, P=3.9\% and P=13\% respectively) the combined probability for these 4 clusters is $9e^{-5}$, or a 
3.7$\sigma$ discrepancy. However, as mentioned earlier, there is a very small number of such clusters observed in the whole sky in comparison with the size 
of the Millennium Simulation, and therefore it seems more plausible to reconcile the observations with whole-sky simulations.

%Using their estimates of the cumulative probability distribution $P(>R_E)$ for an agreement between this $R_E$ and 
%the Millennium simulation $\Lambda$CDM, we find a higher probability P=20\% (instead of 7.9\%) of agreement, making Abell 1703 less 
%discrepant than the other mentioned clusters (Abell 1689, CL0024 and RXJ1347) for its virial mass. .As mentioned earlier, considering the inherent biases of 
%this cluster sample towards large concentration values, there is a good agreement of the Einstein radius with predictions from the simulations. 

%__________________________________________________________________
%%%%%%%%%%%%%%%%%%%%%%%%%%%%%%%%%%%%%%%%%%%%%%%%%%%%%%%%%%%%%%%%

%%%%%%%%%%%%%%%%%%%%%%%%%%%%%%%%%%%%%%%%%%%%%%%%%%%%%%%%%%%%%%%%
\section{\label{discussion}Abell 1703: a relaxed, unimodal cluster ?}

Our strong lensing analysis using a simple NFW component for the large scale dark matter distribution, after individual treatment 
of the central cD galaxy and small scale mass distributions associated with the galaxy substructure in the cluster, is able to reproduce 
the large number of multiple images identified in Abell 1703. Even if the overall RMS in the predicted location of the multiple images (1.3\arcsec ) is 
still quite larger than the precision in astrometric measurements with HST (typically 0.1\arcsec ), we find no systematics as a function of the location of 
the multiples, that would argue for the need of secondary dark matter clumps. This was the case in the clusters Abell 2218 \citep{Ardis} and Abell 68 \citep{Richard07} 
which show a strong evidence for {\it bimodality}. 

Furthermore, adding new spectroscopic measurements on 7 multiple sources has removed an identical number of free parameters (i.e., the 
redshifts of the sources) from the analysis presented in L08. However, we found no significant increase in the overall RMS of the 
multiple images compared to this previous work, under the same assumptions.  In addition,  the strongest variations in the best fit parameters compared with the previous 
values presented in L08 are found for the gNFW profile, but both results are compatible within the 3 $\sigma$ confidence level. 
The remaining scatter in the positions of multiple images is very likely due to small-scale substructure in the dark matter distribution, which is not included here 
apart from the visible cluster galaxies. One way to increase the complexity of this model would be to estimate small-scale deviations from the smooth mass distribution, 
using adaptive-grid based models (Jullo et al. 2009, submitted).

Finally, the best fit parameters of our strong-lensing analysis have shown to be consistent with weak-lensing measurements. Our best fit concentration parameter 
$c_{200}=4.72^{+0.43}_{-0.45}$, corresponding to $c_{vir}=5.82^{+0.53}_{-0.56}$ with the second definition based on the virial radius $r_{vir}$,
is somewhat lower than the value derived by \citet{Broadhurst08a}, $c_{vir}=9.92^{+2.39}_{-1.63}$, but marginally compatible at 2.4$\sigma$ 
level. Despite this smaller concentration parameter, the predicted shear profile provides an overall good fit to their observations: the main difference 
is our smaller Einstein radius (which sets the location of the first datapoint in Fig. \ref{shear}, as was done by \citealt{Broadhurst08a}), allowing for a smaller 
slope of the shear profile. We also note that the two shear measurements at large radii (where the signal gets weaker) are slightly lower than the predictions  
(but consistent within their error bars). Together, these differences affect the best fit concentration parameter. The new Einstein radius measurement should be 
more reliable, since it is based on a large number of multiple images at the center. There seems to be a smaller discrepancy 
between our new strong lensing NFW parameters and the weak lensing measurements, as it was the case for Abell 1689 \citep{Broadhurst05,Limousin07}.

Overall, we argue that these results provide further evidence for Abell 1703 to be a relaxed cluster, as was already suggested in L08. Follow-up 
X-ray observations of the cluster core should allow to confirm again this assumption, and also to separate the X-ray gas component from the 
dark matter in the mass distribution when doing the strong-lensing analysis, following the work by \citet{Marusa} in RXJ1347. Another independent 
measurement of the mass profile would be to study the dynamics of cluster galaxies up to large radii, giving another estimate of the virial radius 
and virial mass. Finally, we expect to find a larger sample of similar relaxed clusters from the ongoing LoCuSS survey \citep{locuss} of clusters with lower 
masses (as inferred from their X-ray luminosity) at $z\sim0.2$. By pursuing the same strong lensing analysis on this broader sample, we will 
probe a wider distribution of Einstein radii, ellipticities and concentration parameters with less bias compared to available numerical simulations. 

%______________________________________________________________
%%%%%%%%%%%%%%%%%%%%%%%%%%%%%%%%%%%%%%%%%%%%%%%%%%%%%%%%%%%%%%%%

\begin{acknowledgements}
We thank the anonymous referee for his/her helpful comments, and acknowledge useful discussions with Masamune Oguri, Carlos Frenk and Mark Swinbank. JR acknowledges support from an European union Marie-Curie Fellowship.
ML acknowledges the Agence Nationale de la Recherche for its support, project number 06-BLAN-0067. The Dark Cosmology Center is funded by the Danish National Research Foundation. 
We thank the Danish Centre for Scientific Computing at the University of Copenhagen for providing us generous amount of time on its supercomputing facility. JPK aknowledges the Centre National de la Recherche Scientifique for its support. The authors recognize and acknowledge the very significant cultural role and reverence that the summit
of Mauna Kea has always had within the indigenous Hawaiian community.
We are most fortunate to have the opportunity to conduct observations from this
mountain.

\end{acknowledgements}

%\begin{thebibliography}{}

\bibliography{references}

\begin{thebibliography}{45}
\expandafter\ifx\csname natexlab\endcsname\relax\def\natexlab#1{#1}\fi

\bibitem[{{Adelberger} \& {Steidel}(2000)}]{Adelberger}
{Adelberger}, K.~L. \& {Steidel}, C.~C. 2000, \apj, 544, 218

\bibitem[{{Bolzonella} {et~al.}(2000){Bolzonella}, {Miralles}, \& {Pell{\'
  o}}}]{hyperz}
{Bolzonella}, M., {Miralles}, J.-M., \& {Pell{\' o}}, R. 2000, \aap, 363, 476

\bibitem[{{Bouwens} {et~al.}(2006){Bouwens}, {Illingworth}, {Blakeslee}, \&
  {Franx}}]{Bouwens06}
{Bouwens}, R.~J., {Illingworth}, G.~D., {Blakeslee}, J.~P., \& {Franx}, M.
  2006, \apj, 653, 53

\bibitem[{{Bouwens} {et~al.}(2008){Bouwens}, {Illingworth}, {Franx}, \&
  {Ford}}]{Bouwens08}
{Bouwens}, R.~J., {Illingworth}, G.~D., {Franx}, M., \& {Ford}, H. 2008, \apj,
  686, 230

\bibitem[{{Brada{\v c}} {et~al.}(2008){Brada{\v c}}, {Schrabback}, {Erben},
  {McCourt}, {Million}, {Mantz}, {Allen}, {Blandford}, {Halkola},
  {Hildebrandt}, {Lombardi}, {Marshall}, {Schneider}, {Treu}, \&
  {Kneib}}]{Marusa}
{Brada{\v c}}, M., {Schrabback}, T., {Erben}, T., {et~al.} 2008, \apj, 681, 187

\bibitem[{{Broadhurst} {et~al.}(2005){Broadhurst}, {Ben{\'{\i}}tez}, {Coe},
  {Sharon}, {Zekser}, {White}, {Ford}, {Bouwens}, {Blakeslee}, {Clampin},
  {Cross}, {Franx}, {Frye}, {Hartig}, {Illingworth}, {Infante}, {Menanteau},
  {Meurer}, {Postman}, {Ardila}, {Bartko}, {Brown}, {Burrows}, {Cheng},
  {Feldman}, {Golimowski}, {Goto}, {Gronwall}, {Herranz}, {Holden}, {Homeier},
  {Krist}, {Lesser}, {Martel}, {Miley}, {Rosati}, {Sirianni}, {Sparks},
  {Steindling}, {Tran}, {Tsvetanov}, \& {Zheng}}]{Broadhurst05}
{Broadhurst}, T., {Ben{\'{\i}}tez}, N., {Coe}, D., {et~al.} 2005, \apj, 621, 53

\bibitem[{{Broadhurst} {et~al.}(2008){Broadhurst}, {Umetsu}, {Medezinski},
  {Oguri}, \& {Rephaeli}}]{Broadhurst08a}
{Broadhurst}, T., {Umetsu}, K., {Medezinski}, E., {Oguri}, M., \& {Rephaeli},
  Y. 2008, \apjl, 685, L9

\bibitem[{{Broadhurst} \& {Barkana}(2008)}]{Broadhurst08b}
{Broadhurst}, T.~J. \& {Barkana}, R. 2008, \mnras, 390, 1647

\bibitem[{{Covone} {et~al.}(2006){Covone}, {Kneib}, {Soucail}, {Richard},
  {Jullo}, \& {Ebeling}}]{Covone}
{Covone}, G., {Kneib}, J.-P., {Soucail}, G., {et~al.} 2006, \aap, 456, 409

\bibitem[{{Dow-Hygelund} {et~al.}(2005){Dow-Hygelund}, {Holden}, {Bouwens},
  {van der Wel}, {Illingworth}, {Zirm}, {Franx}, {Rosati}, {Ford}, {van
  Dokkum}, {Stanford}, {Eisenhardt}, \& {Fazio}}]{Dow}
{Dow-Hygelund}, C.~C., {Holden}, B.~P., {Bouwens}, R.~J., {et~al.} 2005, \apjl,
  630, L137

\bibitem[{{Dubinski}(1998)}]{Dubinski}
{Dubinski}, J. 1998, \apj, 502, 141

\bibitem[{{Egami} {et~al.}(2005){Egami}, {Kneib}, {Rieke}, {Ellis}, {Richard},
  {Rigby}, {Papovich}, {Stark}, {Santos}, {Huang}, {Dole}, {Le Floc'h}, \&
  {P{\'e}rez-Gonz{\'a}lez}}]{Egami}
{Egami}, E., {Kneib}, J.-P., {Rieke}, G.~H., {et~al.} 2005, \apjl, 618, L5

\bibitem[{{El{\'{\i}}asd{\'o}ttir} {et~al.}(2007){El{\'{\i}}asd{\'o}ttir},
  {Limousin}, {Richard}, {Hjorth}, {Kneib}, {Natarajan}, {Pedersen}, {Jullo},
  \& {Paraficz}}]{Ardis}
{El{\'{\i}}asd{\'o}ttir}, {\'A}., {Limousin}, M., {Richard}, J., {et~al.} 2007,
  astro-ph/0710.5636

\bibitem[{{Faber} \& {Jackson}(1976)}]{FJ}
{Faber}, S.~M. \& {Jackson}, R.~E. 1976, \apj, 204, 668

\bibitem[{{Faltenbacher} {et~al.}(2005){Faltenbacher}, {Allgood},
  {Gottl{\"o}ber}, {Yepes}, \& {Hoffman}}]{Faltenbacher}
{Faltenbacher}, A., {Allgood}, B., {Gottl{\"o}ber}, S., {Yepes}, G., \&
  {Hoffman}, Y. 2005, \mnras, 362, 1099

\bibitem[{{Gavazzi} {et~al.}(2003){Gavazzi}, {Fort}, {Mellier}, {Pell{\'o}}, \&
  {Dantel-Fort}}]{Gavazzi}
{Gavazzi}, R., {Fort}, B., {Mellier}, Y., {Pell{\'o}}, R., \& {Dantel-Fort}, M.
  2003, \aap, 403, 11

\bibitem[{{Golse} {et~al.}(2002){Golse}, {Kneib}, \& {Soucail}}]{Golse}
{Golse}, G., {Kneib}, J.-P., \& {Soucail}, G. 2002, \aap, 387, 788

\bibitem[{{Hashimoto} {et~al.}(2008){Hashimoto}, {Henry}, \&
  {Boehringer}}]{Hashimoto}
{Hashimoto}, Y., {Henry}, J.~P., \& {Boehringer}, H. 2008, \mnras, 390, 1562

\bibitem[{{Hennawi} {et~al.}(2008){Hennawi}, {Gladders}, {Oguri}, {Dalal},
  {Koester}, {Natarajan}, {Strauss}, {Inada}, {Kayo}, {Lin}, {Lampeitl},
  {Annis}, {Bahcall}, \& {Schneider}}]{Hennawi}
{Hennawi}, J.~F., {Gladders}, M.~D., {Oguri}, M., {et~al.} 2008, \aj, 135, 664

\bibitem[{{Jullo} {et~al.}(2007){Jullo}, {Kneib}, {Limousin},
  {El{\'{\i}}asd{\'o}ttir}, {Marshall}, \& {Verdugo}}]{Jullo}
{Jullo}, E., {Kneib}, J.-P., {Limousin}, M., {et~al.} 2007, New Journal of
  Physics, 9, 447

\bibitem[{{Kashikawa} {et~al.}(2006){Kashikawa}, {Shimasaku}, {Malkan}, {Doi},
  {Matsuda}, {Ouchi}, {Taniguchi}, {Ly}, {Nagao}, {Iye}, {Motohara},
  {Murayama}, {Murozono}, {Nariai}, {Ohta}, {Okamura}, {Sasaki}, {Shioya}, \&
  {Umemura}}]{Kashikawa}
{Kashikawa}, N., {Shimasaku}, K., {Malkan}, M.~A., {et~al.} 2006, \apj, 648, 7

\bibitem[{{Kelson}(2003)}]{Kelson}
{Kelson}, D.~D. 2003, \pasp, 115, 688

\bibitem[{{Kneib} {et~al.}(2004){Kneib}, {Ellis}, {Santos}, \&
  {Richard}}]{Kneib04}
{Kneib}, J.-P., {Ellis}, R.~S., {Santos}, M.~R., \& {Richard}, J. 2004, \apj,
  607, 697

\bibitem[{{Kneib} {et~al.}(1996){Kneib}, {Ellis}, {Smail}, {Couch}, \&
  {Sharples}}]{Kneib96}
{Kneib}, J.-P., {Ellis}, R.~S., {Smail}, I., {Couch}, W.~J., \& {Sharples},
  R.~M. 1996, \apj, 471, 643

\bibitem[{{Le Fevre} {et~al.}(1995){Le Fevre}, {Crampton}, {Lilly}, {Hammer},
  \& {Tresse}}]{lefevre}
{Le Fevre}, O., {Crampton}, D., {Lilly}, S.~J., {Hammer}, F., \& {Tresse}, L.
  1995, \apj, 455, 60

\bibitem[{{Limousin} {et~al.}(2007){Limousin}, {Richard}, {Jullo}, {Kneib},
  {Fort}, {Soucail}, {El{\'{\i}}asd{\'o}ttir}, {Natarajan}, {Ellis}, {Smail},
  {Czoske}, {Smith}, {Hudelot}, {Bardeau}, {Ebeling}, {Egami}, \&
  {Knudsen}}]{Limousin07}
{Limousin}, M., {Richard}, J., {Jullo}, E., {et~al.} 2007, \apj, 668, 643

\bibitem[{{Limousin} {et~al.}(2008){Limousin}, {Richard}, {Kneib}, {Brink},
  {Pell{\'o}}, {Jullo}, {Tu}, {Sommer-Larsen}, {Egami}, {Micha{\l}owski},
  {Cabanac}, \& {Stark}}]{Limousin08}
{Limousin}, M., {Richard}, J., {Kneib}, J.-P., {et~al.} 2008, \aap, 489, 23

\bibitem[{{Link} \& {Pierce}(1998)}]{Link}
{Link}, R. \& {Pierce}, M.~J. 1998, \apj, 502, 63

\bibitem[{{Madau}(1995)}]{Madau}
{Madau}, P. 1995, \apj, 441, 18

\bibitem[{{Medezinski} {et~al.}(2007){Medezinski}, {Broadhurst}, {Umetsu},
  {Coe}, {Ben{\'{\i}}tez}, {Ford}, {Rephaeli}, {Arimoto}, \&
  {Kong}}]{Medezinski}
{Medezinski}, E., {Broadhurst}, T., {Umetsu}, K., {et~al.} 2007, \apj, 663, 717

\bibitem[{{Natarajan} {et~al.}(1998){Natarajan}, {Kneib}, {Smail}, \&
  {Ellis}}]{Natarajan98}
{Natarajan}, P., {Kneib}, J.-P., {Smail}, I., \& {Ellis}, R.~S. 1998, \apj,
  499, 600

\bibitem[{{Navarro} {et~al.}(1997){Navarro}, {Frenk}, \& {White}}]{NFW}
{Navarro}, J.~F., {Frenk}, C.~S., \& {White}, S.~D.~M. 1997, \apj, 490, 493

\bibitem[{{Oguri} \& {Blandford}(2008)}]{Oguri}
{Oguri}, M. \& {Blandford}, R.~D. 2008, MNRAS accepted, astro-ph/0808.0192

\bibitem[{{Oke} {et~al.}(1995){Oke}, {Cohen}, {Carr}, {Cromer}, {Dingizian},
  {Harris}, {Labrecque}, {Lucinio}, {Schaal}, {Epps}, \& {Miller}}]{lris}
{Oke}, J.~B., {Cohen}, J.~G., {Carr}, M., {et~al.} 1995, \pasp, 107, 375

\bibitem[{{Richard} {et~al.}(2007){Richard}, {Kneib}, {Jullo}, {Covone},
  {Limousin}, {Ellis}, {Stark}, {Bundy}, {Czoske}, {Ebeling}, \&
  {Soucail}}]{Richard07}
{Richard}, J., {Kneib}, J.-P., {Jullo}, E., {et~al.} 2007, \apj, 662, 781

\bibitem[{{Richard} {et~al.}(2008){Richard}, {Stark}, {Ellis}, {George},
  {Egami}, {Kneib}, \& {Smith}}]{Richard08}
{Richard}, J., {Stark}, D.~P., {Ellis}, R.~S., {et~al.} 2008, \apj, 685, 705

\bibitem[{{Sand} {et~al.}(2008){Sand}, {Treu}, {Ellis}, {Smith}, \&
  {Kneib}}]{Sand}
{Sand}, D.~J., {Treu}, T., {Ellis}, R.~S., {Smith}, G.~P., \& {Kneib}, J.-P.
  2008, \apj, 674, 711

\bibitem[{{Shimasaku} {et~al.}(2006){Shimasaku}, {Kashikawa}, {Doi}, {Ly},
  {Malkan}, {Matsuda}, {Ouchi}, {Hayashino}, {Iye}, {Motohara}, {Murayama},
  {Nagao}, {Ohta}, {Okamura}, {Sasaki}, {Shioya}, \& {Taniguchi}}]{Shimasaku}
{Shimasaku}, K., {Kashikawa}, N., {Doi}, M., {et~al.} 2006, \pasj, 58, 313

\bibitem[{{Smith} {et~al.}(2005){Smith}, {Kneib}, {Smail}, {Mazzotta},
  {Ebeling}, \& {Czoske}}]{Smith05}
{Smith}, G.~P., {Kneib}, J.-P., {Smail}, I., {et~al.} 2005, \mnras, 359, 417

\bibitem[{{Stanway} {et~al.}(2004){Stanway}, {Bunker}, {McMahon}, {Ellis},
  {Treu}, \& {McCarthy}}]{Stanway04}
{Stanway}, E.~R., {Bunker}, A.~J., {McMahon}, R.~G., {et~al.} 2004, \apj, 607,
  704

\bibitem[{{Stanway} {et~al.}(2005){Stanway}, {McMahon}, \&
  {Bunker}}]{Stanway05}
{Stanway}, E.~R., {McMahon}, R.~G., \& {Bunker}, A.~J. 2005, \mnras, 359, 1184

\bibitem[{{Umetsu} \& {Broadhurst}(2008)}]{Umetsu}
{Umetsu}, K. \& {Broadhurst}, T. 2008, \apj, 684, 177

\bibitem[{{Verdugo} {et~al.}(2007){Verdugo}, {de Diego}, \&
  {Limousin}}]{Verdugo}
{Verdugo}, T., {de Diego}, J.~A., \& {Limousin}, M. 2007, \apj, 664, 702

\bibitem[{{Zhang} {et~al.}(2008){Zhang}, {Finoguenov}, {B{\"o}hringer},
  {Kneib}, {Smith}, {Kneissl}, {Okabe}, \& {Dahle}}]{locuss}
{Zhang}, Y.-Y., {Finoguenov}, A., {B{\"o}hringer}, H., {et~al.} 2008, \aap,
  482, 451

\bibitem[{{Zhao}(1996)}]{Zhao}
{Zhao}, H. 1996, \mnras, 278, 488

\end{thebibliography}

%\end{thebibliography}

%%%%%%%%%%%%%%%%%%%%%%%%%%%%%%%%%%%%%%%%%%%%%%%%%%%%%%%%%%%%%%%%

% TABLES AND FIGURES

%%%%%%%%%%%%%%%%%%%%%%%%%%%%%%%%%%%%%%%%%%%%%%%%%%%%%%%%%%%%%%%%
\begin{table}
\begin{tabular}{llcc}
\hline
Filter  & $\lambda$ ($\mu$m) & Exposure Time (ksec) & Depth (AB)\\
\hline
ACS/F435W & 0.435 & 7.05 & 26.5 \\
ACS/F475W & 0.480 & 5.56 & 27.5\\
ACS/F555W & 0.539 & 5.56 & 27.1 \\
ACS/F625W & 0.635 & 8.49 & 27.6 \\
ACS/F775W & 0.779 & 11.13 & 27.3\\
ACS/F850LP & 0.908 & 17.8 &26.9 \\
NICMOS/F110W & 1.147 & 0.26 & 25.4 \\ 
MOIRCS/H & 1.642 & 18.87 & 25.3\\
\end{tabular}
\caption{\label{images} Summary of the broad-band images used for the photometry. The depth 
is given as the 5$\sigma$ detection limit for a point source.
}
\end{table}

\begin{table*}
\begin{tabular}{lccllrccl}
\hline
ID & $\alpha$ (J2000.0) & $\delta$ (J2000.0) & $z_{spec}$ & $z_{phot}$ & F775W & $\mu$ (mags) & $z_{class}$ & Features\\
\hline 
1.1 & 198.77725 & 51.81934& 0.8889$\pm$0.0003  & 0.965$^{+0.075}_{-0.240}$      & 22.19$\pm$0.01 & 4.78$\pm$0.22 & 4 & $[{\rm OII}]$, $[{\rm NeIII}]$, H$\delta$, H$\gamma$\\ 
3.1 & 198.76696 & 51.83205& 3.277 $\pm$0.002  & 3.35$^{+0.052}_{-0.134}$            & 22.69$\pm$0.02 & 3.40$\pm$0.47  & 4 & $Ly_\alpha$(em), SiIV, SiII\\
3.3 & 198.75824 & 51.82982 & 3.277 $\pm$0.002  & 3.35$^{+0.024}_{-0.036}$           & 25.34$\pm$0.05 & 1.66$\pm$0.22 & 4 & $Ly_\alpha$(em), SiIV, SiII\\
4.2/5.2 & 198.76075 &51.82487 &1.9082 $\pm$0.004 &  2.25$^{+0.216}_{-0.222}$    & 23.72$\pm$0.02 &2.54$\pm$0.03 & 3 & ${\rm CIV}$, SiII, ZnII \\  
6.1 & 198.77984 & 51.82640 & 2.360  $\pm$0.002  & 2.59$^{+0.144}_{-0.159}$          & 23.56$\pm$0.02 & 1.96$\pm$0.04 & 4 & $Ly_\alpha$(em), SiIV, CIV, SiII\\
7.3 & 198.75869 & 51.82814 & 2.983   $\pm$0.002 & 3.20$^{+0.210}_{-0.675}$          & 26.81$\pm$0.10 & 1.87$\pm$0.02 & 9 & $Ly_\alpha$(em)\\
10.1 & 198.78708 & 51.81424& 2.627 $\pm$0.004 &  3.10$^{+0.324}_{-0.162}$          & 21.40$\pm$0.01 & 2.69$\pm$0.65 & 3 &  $Ly_\alpha$(abs), SiIV\\
11.3  & 198.76242 & 51.80954 & 2.627 $\pm$0.004 & 2.70$^{+0.189}_{-0.261}$         & 23.55$\pm$0.02 & 1.73$\pm$0.02 & 3 &  $Ly_\alpha$(abs), SiIV, SiII\\
15.1 & 198.76284 & 51.81246& 2.355 $\pm$0.002 & 2.44$^{+0.171}_{-0.231}$           & 24.89$\pm$0.05 & 2.26$\pm$0.10 & 4 &  $Ly_\alpha$(em), CIV\\
15.3 & 198.78825 & 51.82173 & 2.355 $\pm$0.002 & 2.60$^{+0.150}_{-0.363}$          & 25.44$\pm$0.05 & 1.56$\pm$0.05& 4 &  $Ly_\alpha$(em), SiIV, CIV\\
16.1/16.3 & 198.78851 & 51.82096 & 2.810 $\pm$0.004 & 2.75$^{+0.171}_{-0.165}$ & 25.52$\pm$0.06 & 1.61$\pm$0.02& 3 &  $Ly_\alpha$(abs), FeII \\
\hline
20 & 198.76659 & 51.83626 & 1.279 $\pm$ 0.0005& 1.10$^{+0.125}_{-0.126}$           & 21.86$\pm$0.01 & 1.51$\pm$0.03& 4 &$[{\rm OII}]$\\
21 & 198.76000 & 51.83451 & 1.279 $\pm$ 0.0005& 1.13$^{+0.137}_{-0.103}$           & 22.57$\pm$0.02 & 1.75$\pm$0.02 & 4 & $[{\rm OII}]$\\
23 & 198.75612 & 51.80736 & 5.827 $\pm$ 0.0015 & $6.0^{+0.42}_{-0.38}$                 & 27.20$\pm$0.30 & 1.20$\pm$0.03 & 9 & $Ly_\alpha$(em)\\
gal2 & 198.78811 & 51.84276 & 0.5632 $\pm$ 0.0004 & 0.550$^{+0.123}_{-0.140}$  & 20.49$\pm$0.01 & 0.29$\pm$0.01 & 4 & $[{\rm OII}]$, K,H, MgI, MgII\\
gal4 & 198.76596 & 51.85902 & 3.2269  $\pm$ 0.002 & 3.44$^{+0.09}_{-0.07}$           & 24.16$\pm$0.03& 0.40$\pm$0.02 & 3 & $Ly_\alpha$(abs), SiII, SiIV\\
%gal10 & 198.76000 & 51.85480 & 4.356  $\pm$0.002& 4.26$^{+0.11}_{-0.44}$& 1.54$\pm$0.03 \\
gal11-a & 198.76772 & 51.79678 &  2.2864  $\pm$0.002 & 2.56$^{+0.235}_{-0.252}$& 22.86$\pm$0.01 & 0.85$\pm$0.02 & 4 & $Ly_\alpha$(em), SiII, CIV\\
gal12 & 198.77770 & 51.79462 & 2.290  $\pm$0.002   & 2.32$^{+0.255}_{-0.698}$     & 25.32$\pm$0.08 & 0.83$\pm$0.03 & 4 & $Ly_\alpha$(em), CIV, AlIII\\
gal13-b & 198.77518 & 51.78871 & 1.035  $\pm$0.001& 1.05$^{+0.108}_{-0.112}$    & 23.45$\pm$0.03 & 0.48$\pm$0.01 & 4 & $[{\rm OII}]$\\
gal16 & 198.79701 & 51.77606 & 0.3906  $\pm$0.0005& $-$                                            & & 0.10$\pm$0.01 & 4 & K,H\\
gal17 & 198.79065 & 51.77097 & 1.236  $\pm$0.0005& $-$                                              & &0.27$\pm$0.02  & 4 & $[{\rm OII}]$\\
gal18-b & 198.78816 & 51.77364 & 0.5860  $\pm$0.0005& $-$                                        & &0.19$\pm$0.01 & 4 & $[{\rm OII}]$, MgII, K, H\\
gal19-a & 198.77100 & 51.76841 & 0.2909  $\pm$0.0005& $-$                                        & &0.01$\pm0.01$ & 4 & K,H\\
gal19-b  & 198.77063 & 51.76935 & 0.3622  $\pm$0.0005& $-$                                       & &0.06$\pm0.01$ & 4 & $[{\rm OII}]$, K,H\\
\hline
gal1 & 198.78155 & 51.83971 & 0.1432  $\pm$0.0005 & & & & 4 & $[{\rm OII}]$, H$\beta$, $[{\rm OIII}]$ \\
gal3 & 198.77735 & 51.85023 & 0.087  $\pm$0.0005 & & & & 4 & $[{\rm OII}]$, K, H, H$\beta$, $[{\rm OIII}]$ \\
gal11-b & 198.76774 & 51.79744 & 0.280  $\pm$0.001& & & & 4 & $[{\rm OII}]$, H$\beta$, $[{\rm OIII}]$ \\
gal13-a & 198.77540 & 51.78808 & 0.1762  $\pm$0.0005& & & & 4 & K,H\\
gal14 &198.78653 & 51.78574 & 0.1011  $\pm$0.0003& & & & 4 & $[{\rm OII}]$, H$\beta$, $[{\rm OIII}]$\\
gal18-a &  198.78753 & 51.77299 & 0.2705  $\pm$0.0005 & & & & 4 & K,H\\
gal20 & 198.79633 & 51.76602 & 0.2766  $\pm$0.0005 & & & & 4 & K,H\\
gal21 & 198.78704 & 51.76265 & 0.2781  $\pm$0.0005 & & & & 4 & K,H\\
gal22-a & 198.77246 & 51.75876 & 0.2410  $\pm$0.0005& & & & 4 & $[{\rm OII}]$, K,H\\
gal22-b & 198.77236 & 51.76022 & 0.2766  $\pm$0.0005& & & & 4 & K,H\\
gal23 & 198.75441 & 51.86258 & 0.1765  $\pm$0.0004& & & & 4 & $[{\rm OII}]$, K, H, H$\beta$, $[{\rm OIII}]$ \\
gal24 & 198.75110 & 51.86508 & 0.2683  $\pm$0.0005 & & & & 4 & K,H\\
\end{tabular}
\caption{\label{zspec}Overview of the sources included in the LRIS mask. From left to right: identification, astrometric position, spectroscopic 
redshift, photometric redshift estimate, ACS-F775W band photometry when available, magnification in magnitudes, redshift quality class (see Sect. \ref{zmeasure}), main spectroscopic features.
The mentions (abs) and (em) refer to a Ly$_\alpha$ line in absorption or in emission, respectively.
The first part of the table corresponds to multiply-imaged system, the second part to 
background sources, the third part to cluster members and foreground sources.}
\end{table*}

\begin{table*}
\begin{tabular}{lcclllr}
\hline
ID & $\alpha$ (J2000.0) & $\delta$ (J2000.0) & $z_{model}$ & $z_{phot}$ & F775W & $\mu$ (mags)\\
\hline 
\multicolumn{3}{l}{New multiple images} \\
\hline 
10.4 & 198.77173 & 51.81906 & $z_{spec}=2.627$ & $-$ & 24.77$\pm$0.08 & 1.55$\pm$0.17 \\
\hline 
12.1 & 198.77456 & 51.829946 & {3.31$\pm$0.16} & 3.33$^{+0.47}_{-0.32}$     & 26.68$\pm$0.10 & 1.93$\pm$0.07\\
12.2 & 198.76568 & 51.828658 & & 3.35$^{+0.22}_{-0.31}$                                    & 24.47$\pm$0.04 & 1.87$\pm$0.16 \\
12.3 & 198.75867 & 51.826482 & & 3.16$^{+0.24}_{-0.52}$                                    & 26.33$\pm$0.08 & 2.17$\pm$0.04 \\
%\hline
13.1 & 198.78238 & 51.821204 & {0.839$\pm$0.007} & 1.15$^{+0.13}_{-0.16}$ & 26.93$\pm$0.11& 1.48$\pm$0.03\\
13.2 & 198.76796 & 51.818374 & & 0.68$^{+0.46}_{-0.12}$                                    & 24.18$\pm$0.04 & 2.42$\pm$0.33\\
13.3 & 198.77204 & 51.814973 & & 1.13$^{+0.08}_{-0.10}$                                    & 26.29$\pm$0.08 & 2.16$\pm$0.19\\
13.4 & 198.76692 & 51.815130 & & 1.07$^{+0.19}_{-0.10}$                                    & 25.12$\pm$0.05 & 2.69$\pm$0.06\\
%\hline
14.1 & 198.78662 & 51.820542 & {1.58$\pm$0.02} & $-$                                         & 25.55$\pm$0.11 & 1.63$\pm$0.02\\
14.2 & 198.76868 & 51.820191 & & $-$                                                                        & 25.40$\pm$0.17 & 1.26$\pm$0.48\\
14.3 & 198.76447 & 51.812192 & & $-$                                                                        & 26.33$\pm$0.16 & 2.13$\pm$0.03\\
14.4 & 198.77540 & 51.811945 & & $-$                                                                        & 26.30$\pm$0.15 & 0.81$\pm$0.10\\
\hline
\multicolumn{3}{l}{Other multiple systems from L08}\\
\hline
2.2 & 198.77156 & 51.81174 & 2.16$^{+0.15}_{?0.09}$        & 2.31$^{+0.46}_{-0.38}$ & 25.29$\pm$0.08 & 4.45$\pm$0.18\\
2.3 & 198.76970 & 51.81186 &                                                    & 2.23$^{+0.19}_{-0.66}$ & 26.15$\pm$0.10 & 4.37$\pm$0.16\\
%\hline
8.1 & 198.77250 & 51.83045 & 2.933$^{+0.004}_{-0.077}$  & 2.80$^{+0.17}_{-0.09}$ & 24.95$\pm$0.06 & 2.46$\pm$0.18\\ 
8.2 & 198.76608 & 51.82949 &                                                    & 2.77$^{+0.21}_{-0.11}$ & 23.06$\pm$0.02 & 2.19$\pm$0.10\\
8.3 & 198.75863 & 51.82740 &                                                    & 2.72$^{+0.20}_{-0.12}$ & 25.07$\pm$0.06 & 2.00$\pm$0.03\\ 
%\hline
9.1 & 198.77176 & 51.83030 &  3.271$^{+0.031}_{-0.124}$ & $-$ & 26.68$\pm$0.15 & 3.09$\pm$0.20\\
9.2 & 198.76690 & 51.82957 &                                                    & 2.995$^{+0.195}_{-0.378}$ & 25.32$\pm$0.08 & 2.48$\pm$0.19\\
9.3 & 198.75813 & 51.82708 &                                                    & 3.00$^{+0.37}_{-0.60}$ & 26.92$\pm$0.12 & 1.98$\pm$0.02\\
\end{tabular}
\caption{\label{newmul} Location and redshift estimate of three new multiply-image systems, as well as the radial feature (10.4) associated with the 
giant arc (Systems 10 and 11). 
From left to right: identification, astrometric position, redshift estimate from the lensing model, photometric redshift estimate, magnification in magnitudes. 
We also report the same values for the three remaining systems from L08 (Systems 2, 8 and 9), used in the mass model but lacking spectroscopic redshifts.
The redshift of 10.4 has a spectroscopic measurement and therefore is not a free parameter in the model. System 14 and image 9.1 are too faint to 
provide a reliable photometric redshift estimate.
}
\end{table*}

\begin{table*}
\begin{tabular}{lcccllrcll}
\hline
Clump & (x) & (y) & e & $\theta$ & r (kpc) & $\alpha$ & c$_{200}$ & $\sigma_0$ (kms$^{-1}$) & $z_{model}$ \\
\hline
NFW & -0.79$\pm$0.08 & 0.91$\pm$0.08 & 0.230$\pm$0.006 & 64.0$\pm$0.3 &  476.9$^{+51.6}_{-42.8}$ &  0.92$^{+0.05}_{-0.04}$ &  4.72$^{+0.43}_{-0.45}$ & $-$ & $-$ \\
cD     & [0.0]  & [0.0]  & [0.19]  & [52] & [25] & $Ð$ &  $Ð$ & 355.7$^{+10.3}_{-12.8}$ & $-$ \\
Galaxy 852 & [19.0] & [54.0] & [0.11]  & [65.5]  & 98.7$\pm$1.8& $Ð$ & $Ð$ & 320.9$\pm$3.5 & $-$ \\
L$_{*}$ elliptical galaxy & $Ð$ &  $Ð$  & $Ð$ & $Ð$ & 68.8$^{+0.2}_{-1.8}$ & $Ð$ & $Ð$ &202$^{+2}_{-3}$ & $-$ \\
\end{tabular}
\caption{\label{model} Best fit parameters used to model the mass distribution. The positions (x) and (y) are in arcsecs relative to the central galaxy 
(R.A.=13:15:05.276, Decl.=+51:49:02.85) and oriented in WCS (north is up, east is left). Orientations $\theta$ are measured in degrees and ellipticities ($e$) of the potentials are given as $1-b/a$, where $a$ and $b$ are the semi-major and semi-minor 
axis of the ellipse, respectively. Values in brackets were not optimized by the model. The radius ($r$) refers to the scale radius $r_s$ in the case of the gNFW profile, and the $r_{cut}$ radius 
in the case of the PIEMD profile (see L08 for details). }
\end{table*}

\clearpage

\end{document}